# Measurement and Analysis of UDP Traffic over Wi-Fi and GPRS

Sumit Maheshwari, K. Vasu, Sudipta Mahapatra, C. S. Kumar*
Department of Electronics and Electrical Communication Engineering
*Department of Mechanical Engineering
Indian Institute of Technology, Kharagpur, India
{sumitece87, vasukanster, sudipta.mahapatra, cs.kumar}@gmail.com

*Abstract*—With the increasing usage of mobile devices to ubiquitously access heterogeneous applications in wireless Internet, the measurement and analysis of Internet traffic has become a key research area. In this paper, we present the results of our measurements for VBR traffic over UDP in 802.11g and GPRS networks. We focus on Inter-Packet Arrival Time (IPRT) and Inter-Packet Transmission Delay (IPTD) and observe that the later has a significant impact on the round trip delay. Numerical parameters for Weibull, Exponential and Normal distribution in order to represent such traffic are also presented.

## I. INTRODUCTION

The transport layer provides a mechanism for the exchange of data between end systems. Transmission Control Protocol (TCP) and User Datagram Protocol (UDP) are two main transport protocols which provide connection-oriented and connectionless services respectively. TCP ensures reliable and ordered data delivery while also introducing processing overhead and bandwidth limitations due to congestion and flow control mechanisms. The lightweight UDP neither provides reliable delivery nor suffers from processing overhead and bandwidth limitations and hence is used in time-sensitive applications because dropping packets is preferable to waiting for delayed packets, which may not be an option in a real-time system like Voice over IP (VoIP), IPTV, video on demand and online gaming.

Although TCP is still the popular protocol in the Internet, with the increasing demand of real-time applications, UDP is gaining popularity [1]. The self-similar nature of Internet traffic [2][10] allows researchers to measure and analyze characteristics of both flow level and packet level traffics which give a key to synthetically generate and use the similar traffic for various applications, which is a time consuming process otherwise.

Wireless Internet traffic traces contain inherent information about user behavior, interaction between users, wireless channel, applications and protocols. Thus, analysis of traces is more important than analytical modeling of wireless protocols alone. Though immensely rewarding, analyzing the statistical characteristics of wireless traffic is difficult partly due to the unfamiliar characteristics of wireless network traffic [3].

Wi-Fi and GPRS are two prime modes of accessing Internet via mobiles under the category of wireless Internet; others being 3G and WiMAX which are still gaining pace in terms of usage and accessibility [7]. Therefore, results of real-time measurement and analysis of traffic traces of former networks can be utilized for better capacity management, congestion avoidance and testing before deployment of future networks.

Real-time applications prefer UDP. VoIP for example uses some standard codec like G.711 etc based upon the bandwidth available. By suitably varying the Inter-Packet Transmission Delay (IPTD) depending upon the codec used, congestion at the network can be significantly reduced even for data applications over UDP.

Packet switched networks have great advantages over circuit switched networks. One is that the bandwidth of the circuit is not limited to a small set of fixed allowed rates and the other is that it supports Variable Bit Rate (VBR) traffic [9]. VBR traffic makes efficient use of available bandwidth and thus its analysis becomes important.

The remaining of this paper is organized as follows. Section 2 introduces the testbed set-up for measurements. Section 3 provides the detailed description of the procedure. In Section 4 we present and discuss numerical results. Finally, Section 5 concludes the paper.

## II. WIRELESS TESTBED SET-UP

The wireless IP QoS testbed is set-up in the CAD/CAM laboratory of Indian Institute of Technology Kharagpur, India as shown in fig. 1. It consists two Access Points (APs), 802.11b/g, and a number of wireless (Mobiles, Laptops etc.) and wired devices (like computers) able to access Internet through Wi-Fi AP, LAN or GPRS network. The last mile that is from Access Point (AP) or Base Station (BS; for GPRS) to mobile devices is Wireless.

The server is Intel(R) core(TM)2 Duo CPU with 1.96GB RAM and 2.79GHz processor running Ubuntu 9.3 Operating System (OS) [4]. The client is Nokia N97 mobile with 128 MB RAM running Symbian OS v9.4, Series 60 rel. 5 and ARM 11 434 MHz processor which is able to access GPRS and Wi-Fi 802.11b/g using Universal Plug and Play (UPnP) technology. The mobile supports Mobile Information Device Profile (MIDP 2.1) and thus J2ME applications are used [5].

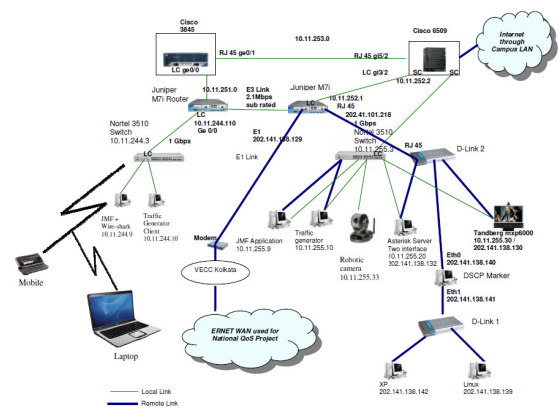

Fig. 1: Wireless IP QoS testbed

The simplified client-server architecture used for testbed used for measurements is shown in fig. 2. BS is shown to depict the GPRS network. Network shown in the picture is different for GPRS and Wi-Fi. Server is set-up in the campus and can be accessed by using IP. All measurements and experiments are done using IPv4.

Java and J2ME based test applications to collect the traces are running on server and client side respectively which are described in the next section.

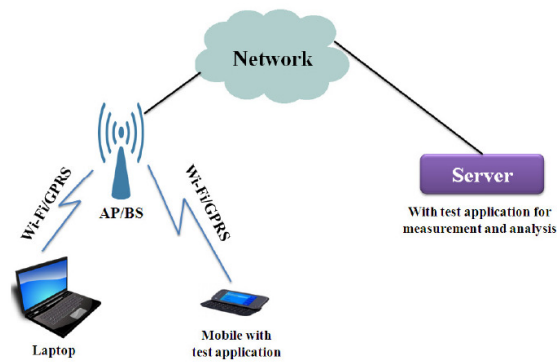

Fig. 2: Client-Server Architecture

## III. MEASUREMENT APPROACH

To accurately measure the send and receive time of packets, we have developed Java based test tools, the architecture of which is shown in fig. 3. The main function of the tools is to time-stamp the packets at both client and server side, store it in the database and retrieve the trace file after completing the experiment.

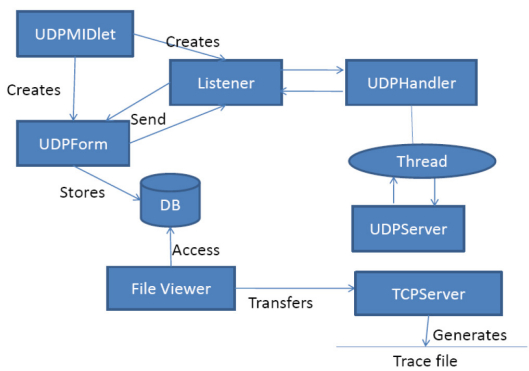

Fig. 3: Test tool architecture

The functions of various components of the tool are as follows:
UDPMIDlet: It runs in mobile client to provide the user interface.
UDPForm: This application is developed to store the actual trace file into the phone database (using RMS).
UDPEchoServer: This is the server application based on UDP to send the Echo reply to the client.

UDPHandler: The actual UDP mechanism is developed in this module.
FileViewerMidlet: It retrieves the trace file from the database and transfers to the TCPServerFile (described below) using socket mechanism.
TCPServerFile: It runs in the server to collect the trace file.

The client generates variable bit rate (VBR) traffic by using random function which sends packet of variable size to the server using Wi-Fi or GPRS network which is selected at the beginning of each experiment. The send time for each packet is logged into the mobile (client). Server upon receiving packets keeps a copy and sends back the original packets to the mobile using the same network as used by the client. Thus, using UDP, which otherwise doesn't give acknowledgement, we are able to calculate round trip delay of a packet which is used in many applications like probing etc. [11]. Client logs the receive time for each packet and keeps the trace file in its memory. The procedure is repeated several times and then the complete trace file is transferred to a system for analysis. The number of consecutive repetitions of the experiment is limited by the fixed buffer size of the mobile device.

VoIP is the specific application of UDP protocol [12] and codec available in the literature suggest that fixed number of constant bit rate (CBR) traffic packets should be sent to the network every unit of time [13]. If VBR traffic is averaged over long period of time gives the feel of CBR traffic due to Long Range Dependence (LRD) [14]. Taking advantage of this fact and applying the features of VoIP codecs to the VBR traffic, we can fix the Inter-Packet Transmission Delay (IPTD) between the consecutive packets sent by the client to avoid congestion at the network as shown in fig 4 and fig 5.

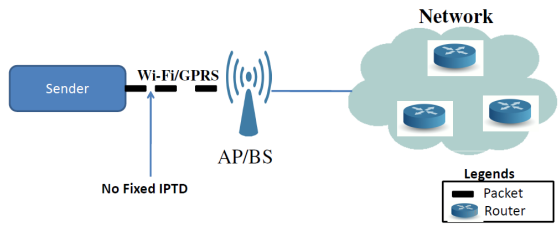
Fig. 4: No fixed Inter Packet Transmission Delay

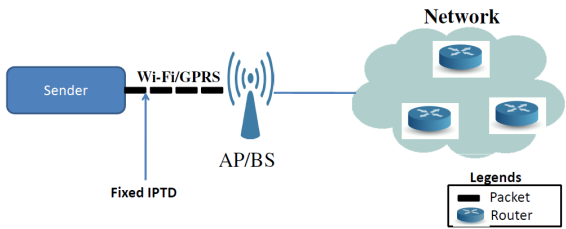
Fig. 5: Fixed Inter Packet Transmission Delay

Taking help of some standard codecs we chose the value of IPTD as 0, 25.31ms and 52.46ms while the choice of the codecs is arbitrary. IPTD=0 implies that there is no delay introduced between the packets.

Experiments are conducted for Wi-Fi and GPRS for all three cases. Round trip time (RTT) is considered as difference between the receive time and send time for the packet at the mobile side [8]. The processing and queuing delay is assumed to be negligibly small. Inter-Packet Arrival Time (IPRT) is the time difference between two successive packets arrived at the sender side. The word sender and mobile are synonymously used throughout the paper.

IV. NUMERICAL RESULTS

The data collected over both the networks i.e. Wi-Fi and GPRS are statistically correlated. The parameters which influence the distribution are Round Trip Time (RTT) and Inter-Packet Arrival Time (IPRT) for different values of IPTD. Data are found to follow some standard distributions whose parameters are obtained using MATLAB. The bin size of the distributions is decided by Sturges' formula or Square root choice for simplicity of the approach. It was observed that the inter packet receive time is between 150ms to 240ms for most of the time in case of IPTD=52.46ms which is well within the limit for a VoIP application where the maximum end to end delay requirement is 300ms. If delay becomes more than that, the listener cannot interpret the message correctly. The RTT is high for larger packet size for IPTD=0 case as compared to IPTD=25 and 52ms.

Since the network is unaware of the application running at the user end and it deals only with the packets, our measured data results can be suitably applied to other real-time applications if the delay is within the tolerable limits. The limited buffer size of the mobile imposed limit on the number of packets it can store in the trace file. Probability density functions are also found for all the data sets obtained (not shown here). A single data set can be can be fit by a number of distributions depending upon the values left while fitting. Few of the histogram results of RTT are shown in fig. 6 and that of IPRT are shown in fig. 7.

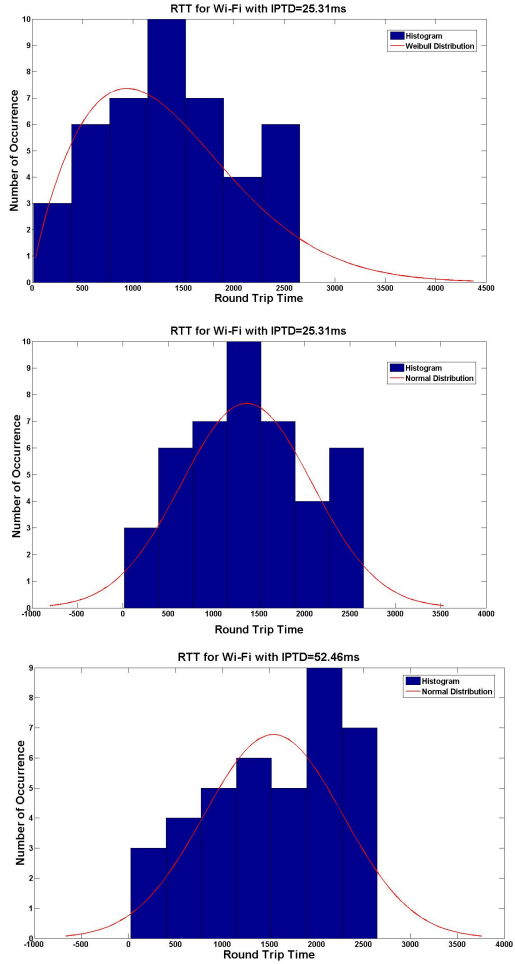

Fig. 7: RTT for Wi-Fi for various IPTD

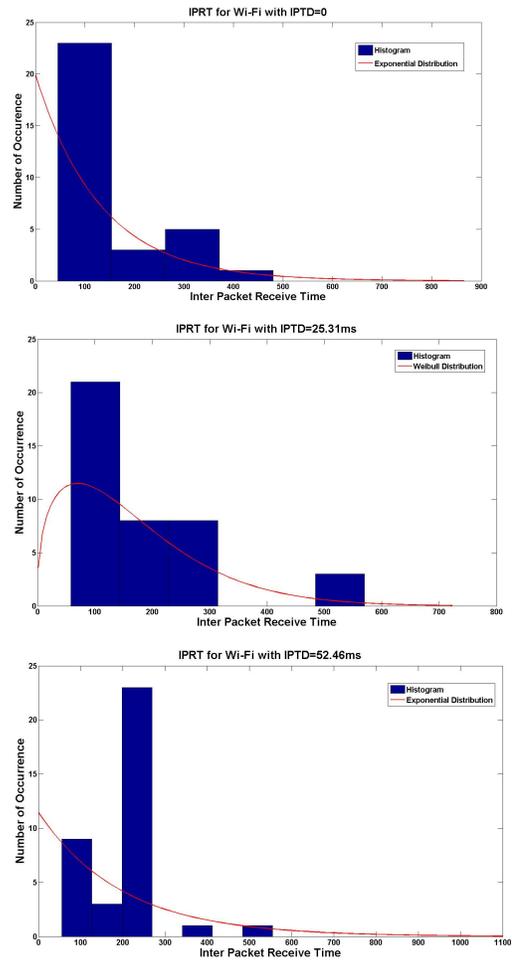

Fig. 8: IPRT for Wi-Fi for various IPTD

RTT shows better performance for Wi-Fi as compared to GPRS as shown in fig. 9. The buffer management in the GPRS network is co-ordinated between the SGSN and the BSC. The rate of the data transmitted from the SGSN to the BSC in the downlink is controlled by the base station subsystem GPRS Protocol (BSSGP). The BSSGP buffers in the SGSN and in the BSC may be considered as one logical BSSGP buffer. After a maximum time determined by the BSSGP buffer setting, data is discarded from the SGSN or the BSC buffer, depending on where data is resided which results in more loss of data in GPRS than in Wi-Fi [6]. With IPTD=0, there a sudden rise in the RTT at a certain packet size which is due to buffering of the packets whereas for IPTD=25ms the RTT increases smoothly. For IPTD=0, the buffering of GPRS is at around 440 bytes packet size while for Wi-Fi it is at 700 bytes. The reason being more number hops in GPRS as compared to Wi-Fi network and the data rate differnces between the two.

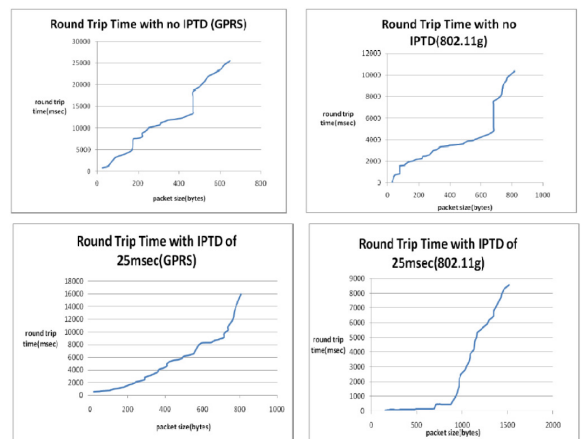

Fig. 9: RTT comparison for Wi-Fi and GPRS for various IPTD

The parameters of various distributions for RTT and IPRT are presented in table 1 and table 2, where the symbols used have usual meanings. RTT gives bestfit for Normal, Exponential and Weibull distributions whereas IPRT shows Exponential and Weibull distributions. All values shown for IPTD and RTT are in milliseconds (ms).

TABLE I. DISTRIBUTION PARAMETERS FOR RTT

| N/W | Round Trip Time | | | |
|---|---|---|---|---|
| | Distribution | IPTD=0 | IPTD=25 | IPTD=52 |
| Wi-Fi | Normal | $\mu=1303.6$ $\sigma^2=639.04$ | $\mu=1364.40$ $\sigma^2=721.9349$ | $\mu=1542.8$ $\sigma^2=737.2472$ |
| | Exponential | $\lambda=0.00077$ | $\lambda=0.00073$ | $\lambda=0.00065$ |
| | Weibull | $\lambda=1.453$ k=0.002 | $\lambda=1.5042$ k=0.0018 | $\lambda=1.7125$ k=0.0021 |
| GPRS | Normal | $\mu=2512.6$ $\sigma^2=1155.01$ | $\mu=4536.8$ $\sigma^2=2209.9$ | $\mu=2747.7$ $\sigma^2=1172.5$ |
| | Exponential | $\lambda=0.0004$ | $\lambda=0.00022$ | $\lambda=0.0036$ |
| | Weibull | $\lambda=2.8396$ k=0.0024 | $\lambda=5.1412$ k=0.0023 | $\lambda=3.0968$ k=0.0026 |

TABLE II. DISTRIBUTION PARAMETERS FOR IPRT

| Parameter | Inter Packet Receive Time (IPRT) | | | |
|---|---|---|---|---|
| N/W | Distribution | IPTD=0 | IPTD=25 | IPTD=52 |
| Wi-Fi | Exponential | $\lambda=0.00108$ | $\lambda=0.00607$ | $\lambda=0.0051$ |
| | Weibull | $\lambda=229.7961$ k=0.5069 | $\lambda=181.8304$ k=1.3682 | $\lambda=223.0583$ k=2.1445 |
| GPRS | Exponential | $\lambda=0.0052$ | $\lambda=0.0024$ | $\lambda=0.0035$ |
| | Weibull | $\lambda=219.3557$ k=1.9359 | $\lambda=471.6538$ k=2.7190 | $\lambda=318.8345$ k=2.8980 |

## V. CONCLUSION

Although TCP is widely accepted transport protocol for most of the Internet applications, UDP is showing growth because of real-time applications' need of fast delivery. The network congestion can be avoided by making best use of UDP by introducing transmission delay between the packets. For the available bandwidth and hence the data rate, the suitable value of IPTD can be chosen which has a significance impact on RTT and IPRT. This paper showed the results of measurements carried out in a wireless testbed for UDP traffic over Wi-Fi and GPRS by using test tools developed specifically for the measurement purpose. The analysis shows that UDP traffic can be imitated by using Exponential, Normal and Weibull distributions. Distribution parameters are also presented to be used by research community.

## ACKNOWLEDGEMENT

This work was carried out under the Vodafone Essar Research Project, IIT Kharagpur, India.